\def\kb{{\bf k}}
\def\qb{{\bf q}}
\def\pb{{\bf p}}
\def\sb{{\bf s}}

\def\rb{{\bf r}}
\def\vb{{\bf v}}

\def\Im{{\rm Im}}
\def\Re{{\rm Re}}
\def\Z3{Z_1^3}

\hfuzz=\maxdimen
\documentstyle[preprint,eqsecnum,prb,aps]{revtex}

\begin{document}
\draft
\title{Quadratic response theory for the interaction of charged particles 
with an electron gas}
\author{J. M. Pitarke and I. Campillo
} \address{Materia Kondentsatuaren Fisika Saila, Zientzi Falkultatea,
Euskal Herriko  Unibertsitatea,\\ 644 Posta kutxatila, 48080 Bilbo, Basque Country,
Spain
}
\date\today
\maketitle

\begin{abstract}
A survey is presented of the theoretical status of quadratic response theories
for the understanding of nonlinear aspects in the interaction of charged
particles with matter. In the frame of the many-body perturbation
theory we study the interaction of charged particles with the electron gas,
within the random-phase approximation (RPA). In particular, nonlinear corrections
to the stopping power of an electron gas for ions are analyzed, and special
emphasis is made on the contribution to the stopping power coming from the
excitation of single and double plasmons. Double plasmon mean free paths of swift
electrons passing through an electron gas are also discussed.
\end{abstract}
\narrowtext

\section{INTRODUCTION}
\label{sec:level1}

A quantitative description of the interaction of charged particles with matter
is of basic importance in many different theoretical and applied
areas\cite{echenique1}. When an ion penetrates condensed matter it causes
changes in the charge state of the ion, electrons may be stripped from the ion
or captured from electronic states of the solid, dynamic screening by valence
electrons originates a wake of electron density fluctuations, and the ion may
lose energy to the medium through different types of elastic and inelastic
collision processes. When a swift electron travels in a solid it may also lose
energy to the medium. While at relativistic velocities radiative losses may
become important, for incident charged particles in the non-relativistic regime
the significant energy losses appear as a consequence of electron-electron
interactions giving rise to the generation of electron-hole pairs, collective
oscillations, and inner-shell excitations and ionizations.

Since the pioneering works of Bohm and Pines\cite{bohm} the response of
conduction electrons in metals to external charged particles has been
represented within the electron gas model, by replacing the ionic lattice by a
homogeneous background which serves to provide neutrality to the system. The
screening properties of a system of interacting electrons are determined,
within linear response theory, by the wavevector and frequency
dependent longitudinal dielectric function $\epsilon_{\qb,\omega}$. In the
self-consistent field, or random-phase, approximation, the dielectric
function of an electron gas was first derived by Lindhard\cite{lindhard}, and,
subsequently, a number of workers have given alternative expressions for
$\epsilon_{\qb,\omega}$, incorporating various many-body higher order
local-field corrections\cite{hubbard,singwi} and band
effects\cite{adler,crawford}. The effect of dissipative processes occurring in a
real metal and conversion of plasmons into multiple electron-hole pairs may be
allowed for in an approximate way by including a damping coefficient in the
dielectric function\cite{mermin}.

Nevertheless, the validity of linear response theory, which treats the
perturbing potential to lowest order, is not obvious a priori.
Although lowest-order perturbation theory leads to energy losses that are
proportional to the square of the projectile charge\cite{bethe}, $Z_1e$, from
measurements on positive and negative pions\cite{barkas} and, also, on protons and
antiprotons\cite{andersen} it is known that the energy loss exhibits a dependence
on the sign of the charge\cite{ashley1}$^-$\cite{pitarke1}. On the other
hand, experimentally observed nonlinear double plasmon
excitations\cite{spence,schatt} cannot be described within linear response
theory\cite{ashley2,pitarke2}, and nonlinearities may also play an important
role on the electronic wake generated by moving ions in an electron
gas\cite{arnau,pitarke3}. Finally, lowest order perturbation theory breaks
down when the projectile is capable of carrying bound electrons with
it\cite{echenique1}.

The first full nonlinear calculation of the electronic stopping power of an
electron gas was performed by Echenique et al\cite{echenique2}, in the
low-velocity limit. They used a scattering theory approach to the stopping power
and the scattering cross sections were calculated for a statically screened
potential which was determined self-consistently by using density-functional
theory. These static screening calculations have recently been extended to
velocities approaching the Fermi velocity\cite{zaremba}. Alternatively, in the
case of incident ions a theoretical effective charge can be
associated\cite{brandt}, and nonlinearities can be investigated, within
quadratic response theory, extending, therefore, the range of linear
response theory and providing results for arbitrary velocity. A
quadratic response theory of the energy loss of charged particles in an
electron gas has recently been carried out\cite{pitarke4}, by following a
diagrammatic analysis of many-body interactions between a moving charge and
the electron gas.

In this paper we present a survey of the theoretical status of investigations
carried out within quadratic response theory for the understanding of
nonlinear aspects in the interaction of charged particles with an electron gas.
We present general procedures to calculate, within many-body perturbation
theory, double plasmon excitation probabilities, $Z_1^3$ contributions to the
stopping power of an electron gas for ions and the nonlinear wake potential
generated by moving ions in an electron gas. We focus on the contribution
to the stopping power coming from the excitation of single and double plasmons.

Unless otherwise is stated, atomic units are used throughout
($\hbar=m_e=e^2=1$).

\section{THEORY}
\label{sec:level1}

We consider a probe of charge $Z_1$ interacting with a many-particle system.
The excitation of eigenmodes of the target together with the reaction of the
probe to these excitations can be described by the self-energy of the probe.
For an incoming particle in a state  $\phi_0$ of energy $p^0$ one
writes\cite{hedin}: 
\begin{equation}
\Sigma_0=\int d^3\rb\int d^3\rb'\phi_0^*(\rb)\Sigma(\rb,\rb',p^0)
\phi_0(\rb'),
\end{equation}
where $\Sigma(\rb,\rb',p^0)$ represents the non-local self-energy.

The real part of $\Sigma_0$ gives us the real energy shift due to the
interaction with the medium, and the imaginary part is well-known to be
directly related to the damping rate experienced by the particle as a
consequence of the interaction with real excitations of the target:
\begin{equation}
\gamma=-2\Im\Sigma_0.
\end{equation}

We take the target to be described by an isotropic homogeneous assembly of
electrons immersed in a uniform background of positive charge and volume
$\Omega$, and we use, therefore, plane waves to describe the incident particle
states. Consequently,
\begin{equation}
\gamma=-2\Im\Sigma_p,
\end{equation}
where $\Sigma_p$ represents the Fourier transform of
$\Sigma(\rb,\rb',p^0)$, $p=(\pb,p^0)$, and $\pb$ is the momentum of the
probe.

The self-energy, $\Sigma_p$, can be calculated in the so-called
GW approximation\cite{quinn,ritchie}:
\begin{equation}
\Sigma_p={\rm i} Z_1^2\int{d^4q\over(2\pi)^4}G_{p-q}W_q,
\end{equation}
where
$G_k$ and $W_q$ represent Fourier transforms of the Green function for the
probe and the time-ordered screened interaction, respectively. In
applying this formula  we  replace $G_k$ by the zero order
approximation; for electrons ($Z_1=-1$)\cite{fetter}:
\begin{equation} 
G_k^0={1-n_{\bf k}\over k^0-\omega_{\bf k}+{\rm i}\eta}+{n_{\bf k}\over
k^0-\omega_{\bf k}-{\rm i}\eta}, \label{13} 
\end{equation} 
where $\omega_\kb=\kb^2/2$, $\eta$ is a positive
infinitesimal, and $n_\kb$ represents the occupation number, which at a
temperature of $T=0K$ is  
\begin{equation}
n_\kb=\theta(k_F-|\kb|),
\end{equation}
$k_F$ being the Fermi momentum and $\theta(x)$,
the Heaviside function.

The dynamically screened interaction, $W_q$, can be represented as follows:
\begin{equation}
W_q=\epsilon_q^{-1}v_\qb,
\end{equation}
where $v_\qb$ represents the Fourier transform of the bare Coulomb interaction:
\begin{equation}
v_\qb={4\pi e^2\over\qb^2},
\end{equation}
and $\epsilon_q$ is the dielectric function, which is related to the
density-density response function, $\chi_q$, by:
\begin{equation}
\epsilon_q^{-1}=1+v_\qb\chi_q.
\end{equation}

Now, introduction of Eqs. (2.5) and (2.7) into Eq. (2.4), and Eq. (2.4) into Eq.
(2.3) gives the following result for the damping rate of incident electrons
with energy above the Fermi level:
\begin{equation}
\gamma=\sum_\qb\int_0^\infty{d q^0\over 2\pi}P_q,
\end{equation}
where $P_q$ represents the probability of transferring four-momentum
$q=({\bf q},q^0)$ to the electron gas:
\begin{equation}
P_q=-{4\pi\over\Omega}Z_1^2\,\Im W_q\,\delta(q^0-p^0+
\omega_{\pb-\qb})\theta(\omega_{\pb-\qb}-E_F).
\end{equation}
The delta function in this expression appears as a consequence of energy
conservation, and the step function, $\theta(\omega_{\pb-\qb}-E_F)$, ensures that
no electrons lose enough energy to fall below the Fermi level. 

When the probe is
not an electron  the occupation number of Eq. (2.6) is zero, i.e., we need not
take account of the fact that the incident electron cannot make transitions
to occupied states in the Fermi sea:
\begin{equation}
P_q=-{4\pi\over\Omega}Z_1^2\,\Im W_q\,\delta(q^0-p^0+
\omega_{\pb-\qb}),
\end{equation}
and if the probe has mass $M>>1$, then recoil can be
neglected in the argument of the delta function to give:
\begin{equation}
P_q=-{4\pi\over\Omega}Z_1^2\,\Im W_q\,\delta(q^0-
\qb\cdot\vb),
\end{equation}
where $\vb$ represents the velocity of the incoming particle.

The inverse mean free path of the probe is easily obtained as follows:
\begin{equation}
\lambda^{-1}={1\over v}\sum_\qb\int_0^\infty{d q^0\over 2\pi}P_q,
\end{equation}
and the stopping power of the target for the probe is obtained as the energy
loss per unit path length of the projectile, after multiplying the probability
$P_q$ by the energy transfer $q^0$:
\begin{equation}
-{dE\over dx}={1\over v}\sum_\qb\int_0^\infty{d q^0\over 2\pi}q^0P_q.
\end{equation}

In the so-called time-dependent Hartree, or random-phase, approximation the
exact linear response function to a screened charge is replaced by the response
function of the non-interacting electron gas:
\begin{equation}
\chi_q^0=-2{\rm i}\int{d^4k\over(2\pi)^4}G_k^0G_{k+q}^0,\label{32}
\end{equation}
thus replacing the linear response function to an external charge,
$\chi_q$, by
\begin{equation}
\chi_q^{RPA}=\chi_q^0+\chi_q^0v_\qb\chi_q^{RPA}.
\end{equation}

Within this approximation the self-energy of Eq. (2.4) can be represented
diagrammatically as in Fig. 1, and cutting the diagrams of this figure through
the two-electron lines in all the bubbles would lead to the open diagrammatic
representation of scattering amplitudes shown in Ref.\onlinecite{pitarke4}. In
particular, if $M>>1$ the incident particle can be treated as a prescribed
source of energy and momentum and one finds\cite{pitarke4}:
\begin{eqnarray} 
S_{f,i}={2\pi{\rm i}\over\Omega}Z_1\int{\rm
d}^4q\delta^4(q+s-p)W^{RPA}_q\delta(q^0-
\qb\cdot\vb),\label{eq:8} 
\end{eqnarray}
where $s=(\sb,s^0)$, $p=(\pb,p^0)$, and $W^{RPA}_q$ represents the random-phase
approximation to the screened interaction of Eq. (2.7).

Then, the probability of transferring four-momentum $q$ to a free-electron gas by
moving a particle from inside the Fermi sea ($|{\bf s}|<q_F$) to
outside ($|{\bf p}|>q_F$), thus creating an electron-hole pair, is derived
from the square of the matrix element $S_{f,i}$:
\begin{equation} 
P_q=2\sum_{{\bf s}}n_{\bf s}\sum_{{\bf p}}(1-n_{\bf
p}) {\left|S_{fi}\right|^2}\delta^4_{q,p-s}, \label{401}  
\end{equation}
where $\delta^4_{q,q'}$ is the symmetric Kronecker
$\delta$ symbol, and
introduction of Eq. (2.18) into Eq. (2.19) gives exactly the result of Eq.
(2.13) found by the self-energy method.

It is obvious at this point that double plasmon excitations cannot be described
within the GW-RPA approximation to the self-energy, represented
diagrammatically in Fig. 1; double excitations can only be described,
within the GW approximation, with inclusion in the screened interaction of
dynamic local-field corrections. On the other hand, the study of
$Z_1^3$ effects in the stopping power of an electron gas for ions and, also, the
study of non-linearities in the wake generated by moving ions in an electron gas
require going beyond the so-called GW approximation. The main ingredient in
the investigation of both double plasmon excitations and $Z_1^3$ effects is the
symmetrized quadratic response function of the non-interacting electron
gas:
\begin{equation}
M_{q,q_1}={\rm i}\int{d^4{k}\over
(2\pi)^4}G_k^0G_{k+q}^0\left[G_{k+q_1}^0+G_{k+q-q_1}^0\right],\label{33}
\end{equation}
which gives account of the quadratic response of the system to a given
charge. The real part of this three-point function was first evaluated by Cenni
et al\cite{cenni}, explicit expressions for the imaginary part in terms of a
sum over hole and particle states have been presented
recently\cite{pitarke1,pitarke4}, and an extension to imaginary frequencies has
also been given\cite{ashcroft}.

\subsection{Double excitation probabilities}

Treating the probe as an external source of energy and momentum, the matrix
element corresponding to the process of carrying the system from an initial
state $a_{i_1}^+a_{i_2}^+|\Phi_0>$ to a final state
$a_{f_1}^+a_{f_2}^+|\Phi_0>$ is 
\begin{equation}
S_{f_1f_2,i_1i_2}={<\Phi_0|a_{f_1}a_{f_2}Sa_{i_1}^+a_{i_2}^+|\Phi_0>\over
<\Phi_0|S|\Phi_0>},\label{14p}
\end{equation}
where $\Phi_0$ is the vacuum state, $a_i$ and $a_i^+$ are annihilation and
creation operators for fermions, respectively, and
$S$ is the scattering matrix. $S$ is obtained as a time-ordered exponential in
terms of the perturbing Hamiltonian and field operators $\Psi(x)$ and
$\Psi^+(x)$ destroying and creating, respectively, a particle at the point $\rb$
at the time $t$. 

Now, one can apply Wick's theorem, we note that only normal
ordered products with four uncontracted field operators contribute, and we
find, up to second order in the probe charge a result that can be represented
diagrammatically as in Fig. 2. Within the random-phase approximation, the
screened interaction, $W_q$, is obtained from Eqs. (2.7), (2.9) and (2.17),
and, accordingly, all self-energy and vertex insertions have been neglected. On
the other hand, exchange processes and, also, ladder contributions have not been
introduced into Eq. (2.21), since they all lead to scattering probabilities that
are of a higher order in the screened interaction.

Finally, the probability for transferring four-momentum $q$ to a free-electron
gas by moving two particles from inside the Fermi sea ($|{\bf s}_1|<q_F$ and
$|{\bf s}_2|<q_F$) to outside ($|{\bf p}_1|<q_F$ and $|{\bf p}_2|<q_F$) is
derived from the square of the matrix element $S_{f_1f_2,i_1i_2}$:
\begin{equation} 
P_q=4\sum_{q_1}\sum_{{\bf s}_1}n_{{\bf s}_1}
\sum_{{\bf s}_2}n_{{\bf s}_2}\sum_{{\bf p}_1}(1-n_{{\bf p}_1})\sum_{{\bf p}_2}(1-n_{{\bf p}_2})
\left|S_{f_1f_2,i_1i_2}\right|^2\delta^4_{q_1,p_1-s_1}
\delta^4_{q-q_1,p_2-s_2}. \label{401p}  
\end{equation}

If the probe were not a heavy particle, then recoil should be introduced into
the argument of the delta function to ensure energy conservation, and, in
particular, if the probe were an electron an step function should also be
introduced to ensure that the probe does not lose enough energy to fall below
the Fermi level. Then the contribution of Eq. (2.22) to the probability that is
proportional to $Z_1^2$, obtained after introduction of the matrix element
$S_{f_1f_2,i_1i_2}$ represented diagrammatically in Fig. 2 into Eq.
(2.22), would coincide with contributions derived from a GW approximation to the
self-energy with inclusion, in the screened interaction, of corresponding
dynamic local-field corrections. 

In particular, the only
$Z_1^2$ contribution to the probability of Eq. (2.22) which might represent
the real excitation of a double plasmon comes from the square of the
scattering amplitude represented by the second diagram of Fig. 2. It is given by
the following expression:
\begin{equation}
P_q={16\pi\over\Omega^2}Z_1^2\,W_q^{-2}\,\sum_{\qb_1}\int_0^{q^0}{dq_1^0\over
2\pi}\,\Im W_{q_1}\,\Im W_{q-q_1}\,|M_{q,q_1}|^2\delta(q^0-p^0+
\omega_{\pb-\qb})\theta(\omega_{\pb-\qb}-E_F).
\end{equation}
Introduction of this probability into Eq. (2.14) gives, after approximating
the linear and quadratic response functions by their low-$q$ limits,
the following high-velocity limit for the $Z_1^2$ contribution to the
inverse mean free path coming from the excitation of a double
plasmon\cite{pitarke2}:
\begin{equation}
\lambda_{2p}^{-1}\approx 0.164{\sqrt{r_s}\over 36\pi v^2}.
\end{equation}
Numerical study shows\cite{campillo} that introduction of the full RPA response
functions gives a result for $\lambda_{2p}^{-1}$ which has, in the high-velocity
limit, the same dependence on $v$ as the approximation of Eq. (2.24), though it
is, for $r_s=2.07$, larger than this approximation by a factor
of $2.16$ .

\subsection{$Z_1^3$ correction to the stopping power for ions}

The stopping power of an electron gas for a probe of charge $Z_1$, mass
$M>>1$ and velocity $\vb$ is obtained after introduction of the probability
$P_q$ into Eq. (2.15). Up to third order in the projectile charge:
\begin{equation}
-{dE\over dx}={1\over v}\sum_\qb\int_0^\infty{d q^0\over 2\pi}
q^0\left(P_q^{\rm single}+P_q^{\rm double}\right),
\end{equation}
where $P_q^{\rm single}$ and $P_q^{\rm double}$, probabilities of transferring
four-momentum
$q$ to the electron gas by creating single and double excitations,
respectively, are obtained from Eqs. (2.19) and (2.22), respectively. There,
\begin{eqnarray}
S_{f,i}=&&{2\pi{\rm }\over\Omega}Z_1\int
d^4q\delta^4(q+s-p)\delta(q^0-\qb\cdot\vb)\cr
&&\times\left\{W_q+2\pi Z_1\left[{\rm i} W_{q_1}W_{q-q_1}G_{s+q}^0
+-{\rm i} W_qW_{q_1}W_{q-q_1}M_{q,q_1}\right]\delta(q_1^0-\qb_1\cdot\vb)
\right\},
\end{eqnarray}
which is represented diagrammatically in Fig. 3, and
\begin{eqnarray}
S&&_{f_1f_2,i_1i_2}={2\pi{\rm i}\over\hbar^2\Omega^2}Z_1\int
d^4q_1\delta^4(q_1+s_1-p_1)\int
d^4q\delta^4(q-q_1+s_2-p_2)\delta(q^0-\qb\cdot\vb)\cr
&&\times\left\{\left[W_qW_{q-q_1}\left(G_{s+q}^0+G_{s-q+q_1}^0\right)
-W_qW_{q_1}W_{q-q_1}M_{q,q_1}\right]
+2\pi{\rm i} Z_1W_{q_1}W_{q-q_1}\delta(q_1^0-\qb_1\cdot\vb)\right\},
\end{eqnarray}
represented diagrammatically in Fig. 2. Processes involving
higher-order excitations have not been included.

Now, after introduction of Eqs. (2.26) and (2.27) into Eqs. (2.19) and (2.22),
respectively, we find, up to third order in the ion charge:
\begin{equation}
-{dE\over dx}=\left(-{dE/dx}\right)^{\rm single_1}
+\left(-{dE/dx}\right)^{\rm single_2}
+\left(-{dE/dx}\right)^{\rm double_2},
\end{equation}
where $\left(-{dE/dx}\right)^{\rm single_1}$ represents the $Z_1^2$
contribution to the stopping power coming from single excitations:
\begin{equation}
\left(-{dE/dx}\right)^{\rm single_1}=
-{2\over v}Z_1^2\int{d^3\qb\over (2\pi)^3}\int_0^\infty{d
q^0}q^0\,\Im W_q\,\delta(q^0-\qb\cdot\vb),
\end{equation}
and $\left(-{dE/dx}\right)^{\rm single_2}$ and $\left(-{dE/
dx}\right)^{\rm double_2}$ represent $Z_1^3$ contributions to the stopping
power coming from single and double excitations, respectively:
\begin{eqnarray}
&&(-{dE/dx})^{\rm single_2}=
-{4\over v}Z_1^3\int{d^3\qb\over
(2\pi)^3}\int_0^\infty{dq^0}q^0\delta(q^0-\qb\cdot\vb)\int{d^3\qb_1\over
(2\pi)^3}\int_{-\infty}^\infty{dq_1^0}\delta(q_1^0-\qb_1\cdot\vb)\cr\cr 
&&\times\left[\Im
W_q\,\Re\left(W_{q_1}W_{q-q_1}M_{q,q_1}\right)
+\Re\left(W_q^*W_{q_1}W_{q-q_1}\right)H_{q,q_1}\right]
\end{eqnarray}
and
\begin{eqnarray}
&&(-{dE/dx})^{\rm double_2}=
-{8\over v}Z_1^3\int{d^3\qb\over
(2\pi)^3}\int_0^\infty{dq^0}q^0\delta(q^0-\qb\cdot\vb)\int{d^3\qb_1\over
(2\pi)^3}\int_0^{q^0}{dq_1^0}\delta(q_1^0-\qb_1\cdot\vb)\cr\cr
&&\times\left[\Im
W_{q_1}\,\Im
W_{q-q_1}\Im\left(W_qM_{q,q_1}\right)
+\Im\left(W_qW_{q_1}^*\right)\Im
W_{q-q_1}\left(H_{q_1,q}+H_{q_1,-(q-q_1)}\right)\right],
\end{eqnarray}
with
\begin{eqnarray}
H_{q,q_1}=&&{8\pi^4 P}\int{{\rm d}^3{\bf s}\over
(2\pi)^3}n_{\bf s}\int{{\rm d}^3{\bf p}\over
(2\pi)^3}(1-n_{\bf p})\delta^3({\bf q}-{\bf p}+{\bf s})\cr
&&\times\left[{\delta(q^0+\omega_{\bf s}-\omega_{\bf p})\over
q_1^0+\omega_{\bf s}-\omega_{{\bf s}+{\bf q}_1}}+{\delta(q^0-\omega_{\bf s}+\omega_{\bf p})
\over
-(q^0-q^0_1)+\omega_{\bf s}-\omega_{{\bf s}+{\bf q}-{\bf
q}_1}}\right]+(q_1\rightarrow q-q_1), 
\label{317}
\end{eqnarray}
which is related to the imaginary part of the three-point function of Eq.
(2.20) by\cite{pitarke1,pitarke4}:
\begin{equation}  
\Im M_{q,q_1}=H_{q,q_1}+H_{q_1,q}+H_{(q-q_1),-q_1}. \label{316}
\end{equation}

The contribution to the $Z_1^2$ stopping power coming from double excitations,
which is of higher order in the screened interaction than the contribution of
Eq. (2.29), has not been included in Eq. (2.28). However, contributions of Eqs.
(2.30) and (2.31), which are proportional to $Z_1^3$, are all of the same order
in the screened interaction, they all need, therefore, to be taken into
account, and they all can be derived from the knowledge of the $Z_1^3$
contribution to the self-energy by going beyond the GW approximation. 

It is interesting to notice that
$Z_1^3$ contributions to the stopping power that are proportional to the
product of two imaginary parts of the screened interaction, appearing as a
consequence of both single and double excitations, can be combined, and we,
also, find that contributions to the $Z_1^3$ stopping power that are
proportional to the product of three imaginary parts of the screened
interaction, coming from single and double excitations, cancel out.

Consequently, one finds the following result for the contribution to the
stopping power that is proportional to $Z_1^3$:
\begin{eqnarray}
&&(-{dE/dx})^{(2)}=(-dE/dx)^{\rm single_2}+(-dE/dx)^{\rm
double_2}=\cr
&&-{4\over v}Z_1^3\int{d^3\qb\over
(2\pi)^3}\int_0^\infty{dq^0}q^0\delta(q^0-\qb\cdot\vb)\int{d^3\qb_1\over
(2\pi)^3}\int_{-\infty}^\infty{dq_1^0}\delta(q_1^0-\qb_1\cdot\vb)\cr
&&\times\left[f_1(q,q_1+f_2(q,q_2)+f_3(q,q_1)\right]
\end{eqnarray}
where
\begin{equation}
f_1(q,q_1)=\Im W_q\,\Re W_{q_1}\,\Re W_{q-q_1}\,\Re
M_{q,q_1},\label{407} \end{equation}
\begin{equation}
f_2(q,q_1)=\Re W_q\,\Re W_{q_1}\,\Re W_{q-q_1}\,H_{q,q_1},\label{408}
\end{equation}
and
\begin{equation}
f_3(q,q_1)=-2\Im W_q\,\Im W_{q_1}\,\Re W_{q-q_1}\,H_{q_1,q}.\label{409}
\end{equation}

Contributions to the $Z_1^3$ stopping power of Eq. (2.34) coming from $f_1$ and
$f_2$ of Eqs. (2.35) and (2.36) both appear as a consequence of single
excitations: the first one comes from the cross product between the first and
third diagrams of Fig. 3, and gives, therefore, the contribution from losses
to one-step single excitations generated by the quadratically screened ion
potential , while the second term comes from the cross product between the
first and second diagrams of Fig. 3 and gives the contribution from losses to
two-step single excitations generated by the linearly screened ion potential.
The third term comes from both cross products, and, also, from losses to
double excitations. Contributions coming from single plasmons are included in
both $f_1$ and $f_3$, and contributions coming from the excitation of double
plasmons are only included in $f_3$.

Alternatively, the stopping power of an electron gas can be obtained from the
knowledge of the wake potential induced in the vicinity of the projectile,
as the induced retarding force that the polarization charge distribution exerts
on the projectile itself, and a second-order many body perturbation analysis of
the wake potential\cite{pitarke3} at {\rb}, defined as the mean value of the
interaction between a test unit positive charge at that point and the electron
gas, leads to Eq. (2.34) for the $Z_1^3$ stopping power, as demonstrated in
Ref.\onlinecite{pitarke4}.
 
For high velocities of the probe the electron gas can be considered as if it
were at rest, one can use, therefore, the so-called static electron gas
approximation for both linear and quadratic response functions, and this
results in $f_2$ and $f_3$ giving no contribution to the integral of Eq.
(2.34), i.e., in the high-velocity limit only the contribution to the $Z_1^3$
stopping power that is proportional to only-one imaginary part of the
linearly screened interaction, $W_q$, is different from zero. Furthermore, it
has been shown\cite{pitarke4} that this contribution to the $Z_1^3$ effect can
be approximated, in the high-velocity limit, by :
\begin{equation} 
(-{dE/dx})^{(2)}=Z_1^3{\omega_p^2\over v^2}L_1,
\end{equation}
 where $\omega_p$ represents the
plasma frequency $\omega_p^2=4\pi n$, $n$ being the electron density of
the medium,  and $L_1$ is the $Z_1^3$ correction to the so-called stopping
number:
\begin{equation}
L_1\approx 1.42{\pi\omega_p\over v^3}\ln{2v^2\over 2.13 \omega_p}.
\label{440}
\end{equation}

At high velocities both the wake potential and the stopping power can also be
derived within a quantum hydrodynamical model of the electron gas. In this
model, we expand the nonlinear hydrodynamical equations and find, after
quantization, a result for the Hamiltonian of the electron plasma-heavy ion
system, in terms of the triple vertex interaction between three excitations,
that exactly agrees with the result obtained by Ashley and
Ritchie\cite{ashley2} by following a different procedure. Then, we find second
and third order wake potentials and electronic stopping powers, and, also,
double plasmon excitation probabilities that coincide with plasmon-pole like
approximations to the full RPA results\cite{bergara}.

\section{results}

Contributions to electron inelastic mean free paths coming from single
excitations of the electron plasma have been calculated in the high velocity
limit\cite{bohm}, and, also, in the full RPA\cite{tung}. Fig. 4 shows, as a
solid line, our full RPA results for the double plasmon inverse mean free path
of electrons passing through an electron gas of a density equal to that of
Aluminum, as a function of the velocity, together with the double plasmon
inverse mean free path of positrons (dashed line) and, also, the high-velocity
limit of Eq. (2.24) multiplied by a factor of 2.16 (dotted line):
\begin{equation}
\lambda_{2p}^{-1}\approx 3.13\times 10^{-3}{\sqrt{r_s}\over v^2}.
\end{equation}

At high velocities of the projectile the electron gas can be considered to be
at rest, the effect of the Pauli restriction is, therefore, removed, and the
behaviour of the double plasmon inverse mean free path, as a function of the
velocity, is independent of the particle statistics. On the other hand, it
is interesting to notice that the high-velocity limit of Eq. (3.1) gives a good
account of the full RPA result for both incident electrons and positrons in
a wide range of projectile velocities. In particular, for Aluminum and an
incident electron energy of $40{\rm keV}$ we find from Eq. (3.1) a ratio for the
double relative to the single plasmon of $1.93\times 10^{-3}$, in agreement
with the experiment\cite{schatt}.

Contributions to the stopping power that are proportional to $Z_1^2$ and
$Z_1^3$, as obtained from Eqs. (2.29) and (2.34), are plotted in Fig. 5 by
solid lines, as a function of the velocity, again for $r_s=2.07$. It is
interesting to notice that both
$Z_1^2$ and $Z_1^3$ contributions to the stopping power exhibit a linear
dependence on the velocity up to velocities approaching the stopping maximum;
this linear dependence has, also, been observed for the low-velocity stopping
power when it is calculated to all orders in the probe charge on the basis of
density functional theory\cite{zaremba}. The linear dependence of the $Z_1^3$
correction to the stopping power is, however, a consequence of two competing
effects. First, there is the effect of one-step single excitations generated by
the quadratically screened ion potential, represented by a dashed line in the
same figure, and, then, the effect of two-step single excitations generated by
the linearly screened ion potential, represented by a dashed-dotted line. The
contribution from losses to two-step single excitations, represented by $f_2$
of Eq. (2.36), is very small, at high velocities, when the velocity distribution
of target electrons can be neglected. In this case the static electron gas
approximation can be made, the only non-vanishing contribution to the $Z_1^3$
effect comes, in this approximation, from $f_1$ of Eq. (2.35), i. e., from
losses to one-step single excitations generated by the quadratically screened
ion potential, and one finds that the result obtained in this approximation is
well reproduced by Eq. (2.38), represented in Fig. 5 by a dotted line. This
approximation gives a good account of the full RPA result, even at intermediate
velocities where the velocity of target electrons is not negligible, and this
is, again, a consequence of two competing effects. First, the non-negligible
motion of the electron gas gives rise to a smaller contribution from losses to
one-step single excitations, and this is almost compensated by the non-vanishing
contribution from losses to two-step single excitations. Contributions from
losses to double excitations are small in a wide range of projectile velocities
and they are exactly equal to zero as far as  the electron gas can be
considered to be at rest.

In order to analyze the contribution to the nonlinear stopping power coming
from losses to collective excitations, we first show in Fig. 6 the separate
contributions to the linear term from plasmon excitation and electron-hole pair
excitation by the incident particle. For a momentum transfer that is smaller
than $q_c$, the critical wave vector where the plasmon dispersion enters the
electron-hole pair excitation spectrum, both the plasmon and the electron-hole
pair excitation contribute to the energy loss, though contributions from losses
from electron-hole pair excitations are very small. For $q>q_c$, however, only
the excitation of electron-hole pairs contributes. Total contributions to the
$Z_1^2$ stopping power coming from
$q<q_c$ and
$q>q_c$ are shown in Fig. 7, and contributions coming from
$q<\sqrt{2\omega_p}$ and $q>\sqrt{2\omega_p}$ are plotted in Fig. 8,
$\sqrt{2\omega_p}$ being the low-density limit of $q_c$. Contributions to
the stopping power coming from losses to plasmons is, therefore, smaller than
contributions from losses to electron-hole pairs, especially at high
electron-densities, though there is, at high velocities, exact equipartition of
the energy loss corresponding to momentum transfers larger than and smaller
than $\sqrt{2\omega_p}$. This equipartition rule appears
straightforwardly in the electron gas at rest approximation, and it has been
formulated, for an electron gas not at rest, by Lindhard and
Winther\cite{winther}. This equipartition  is, also, found to be exact, in the
high velocity limit, by using Coulomb scattering of independent electrons with
$q_{min}=\omega_p/v$ or by assuming that independent electrons are scattered by
a velocity dependent Yukawa potential with screening length proportional to
$\omega_p/v$.

As far as the $Z_1^3$ stopping power is concerned, we have split the
contributions to $f_1$ from losses to single plasmons and single electron-hole
pairs, and we have found the result shown in Fig. 9 by dashed and dotted
lines, respectively. On the other hand, all contributions to $f_2$,
represented in this figure by a dashed-dotted line, come from losses to
electron-hole pairs. Thus, it is obvious from this figure that
contributions to the
$Z_1^3$ effect coming from losses to plasmons is small, showing that nonlinear
corrections to losses from single plasmons are not important, and that
collective exitations appear to be well described by linearly
screened ion potentials. The equipartition rule, valid within first order
perturbation theory and/or a linear response theory of the electron gas,
cannot be extended, therefore, to higher orders in the external perturbation.

Finally, in order to account approximately for the $Z_1^3$ effect coming from
both the conduction band and the inner-shells, a local plasma
approximation has been used, by assuming that a local Fermi energy can be
attributed to each element of the solid, and experimental differences between
the stopping power of silicon for protons and antiprotons have been
successfully explained in this way\cite{pitarke4}. 

\section{Conclusions}

In conclusion, we have developed a quadratic response theory for the
understanding of nonlinear aspects in the interaction of charged particles
with matter. In the frame of many-body perturbation theory, we have studied the
interaction of charged particles with the electron gas, within the
random-phase-approximation, and in particular, the nonlinear wake
potential generated by moving ions in matter, the $Z_1^3$ correction to the
stopping power, and processes involved in multiple excitations of
electron-hole pairs and plasmons.

Double plasmon mean free paths for incident electrons and positrons, and, also,
second order contributions to the stopping power coming from the excitation of
single and double plasmons have been evaluated, for the first time, in the full
RPA, as a function of the velocity of the projectile.

Our results for the $Z_1^3$ correction to the stopping power show that for
velocities smaller than the Fermi velocity the stopping power is, up to third
order in the ion charge, a linear function of the projectile velocity. We have
presented, for the high-velocity limit, a formula that gives a good account of
the full RPA result in a wide range of projectile velocities, and our theory
gives good agreement with the experiment. We have also separated the
contributions to the stopping power coming from losses to plasmon generation,
and we have found that collective excitations are well described by linearly
screened ion potentials.

A nonlinear quantum hydrodynamical model of the electron gas has also been
developed\cite{bergara}. It has been demonstrated that double plasmon
excitation probabilities and the second order wake potential and stopping
power coincide, within this model, with a plasmon-pole like approximation to
our full RPA scheme, and an extension of this model to study the bounded
electron gas is now in progress\cite{bergara1}.

Full calculations of second order contributions to the wake potential and the
induced electron density, within the RPA, for different values of the velocity
of the projectile and the electron density of the medium will be published
elsewhere\cite{bergara2}.

An analysis of the differences between a self-energy approach to the $Z_1^3$
correction to the stopping power and the open diagrammatical approach
presented here, and, also, investigations of the $Z_1^3$ stopping power for
incident electrons and positrons are now in progress.

\acknowledgments

The authors gratefully acknowledge discussions with P. M. Echenique
and R. H. Ritchie. We wish also to acknowledge the support of the University
of the Basque Country, the Basque Unibertsitate eta Ikerketa Saila, and the
Spanish Comisi\'on Asesora, Cient\'\i fica y T\'ecnica (CAICYT).


\begin{figure}
\caption{GW-RPA approximation to the self-energy.}
\end{figure}



\begin{figure}
\caption{Diagrammatic representation, up to second order in the ion charge, of
the RPA $S_{f_1f_2,i_1i_2}$ scattering amplitude. Solid
internal lines in the first and second diagrams are zero-order propagators, and
the triple internal vertex in the second diagram represents the quadratic
density response function of the non-interacting electron gas. All vertex and
self-energy insertions have been neglected, as well as ladder contributions.}
\end{figure}



\begin{figure}
\caption{Diagrammatic representation of the matrix element of Eq. (2.26), as
obtained within the RPA.}
\end{figure}

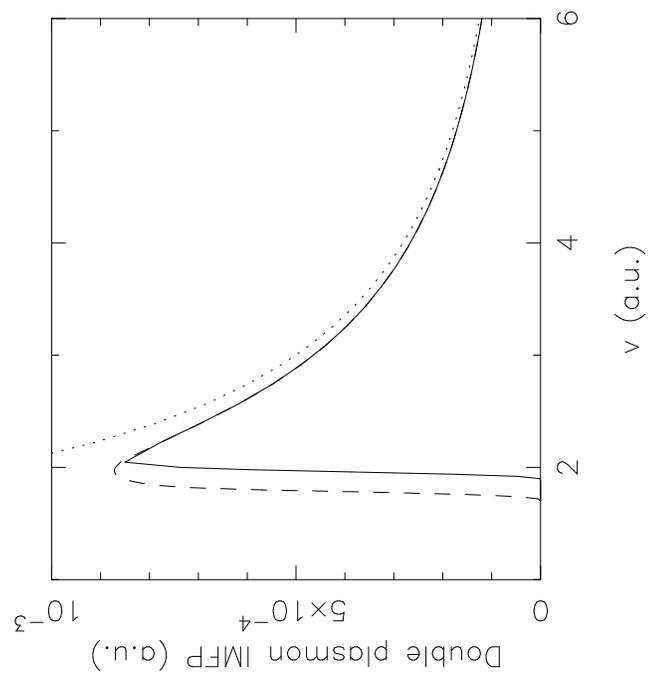
\begin{figure}
\caption{Full RPA double plasmon inverse mean free path of electrons passing
through an electron gas of a density equal to that of Aluminum ($r_s=2.07$),
as a function of the velocity (solid line). The dashed line
represents the double plasmon inverse mean free path of positrons, and the
dotted line, the high-velocity limit of Eq. (2.24).}
\end{figure}

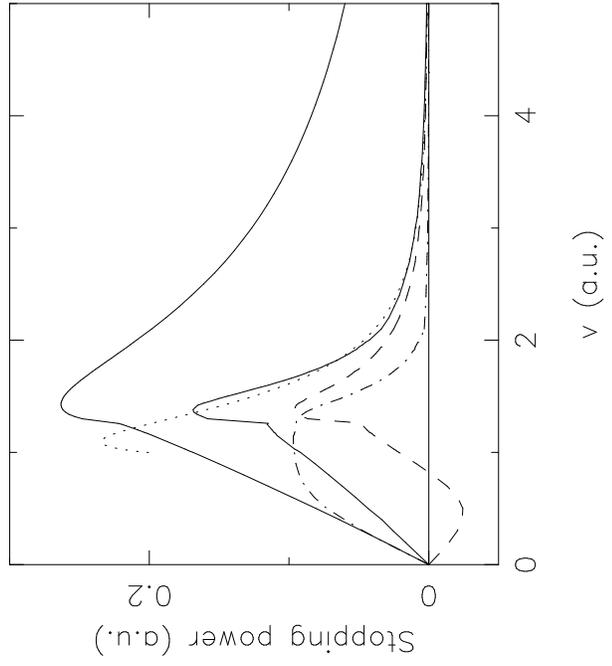
\begin{figure}
\caption{Full RPA $Z_1^2$ and $Z_1^3$ contributions to the stopping power
calculated from Eqs. (2.29) and (2.34), respectively, for $Z_1=1$ and
$r_s=2.07$, as a function of the velocity of the projectile. Dashed and
dashed-dotted lines represent $Z_1^3$ contributions from $f_1$ and $f_2$ of
Eqs. (2.35) and (2.36), respectively. The dotted line represents the
high-velocity limit of Eq. (2.38).}
\end{figure}

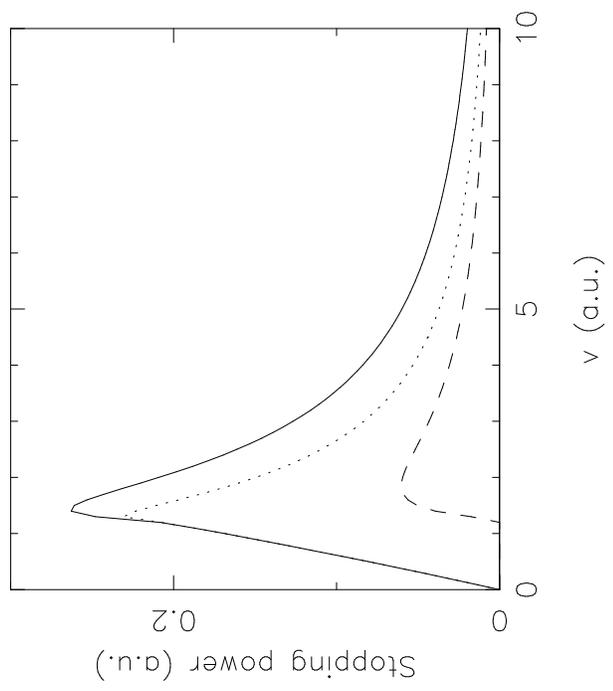
\begin{figure}
\caption{Full RPA $Z_1^2$ contribution to the stopping power (solid line),
versus velocity. Dashed and dotted lines represent contributions from plasmon
and single electron-hole pair excitations, respectively.}
\end{figure}

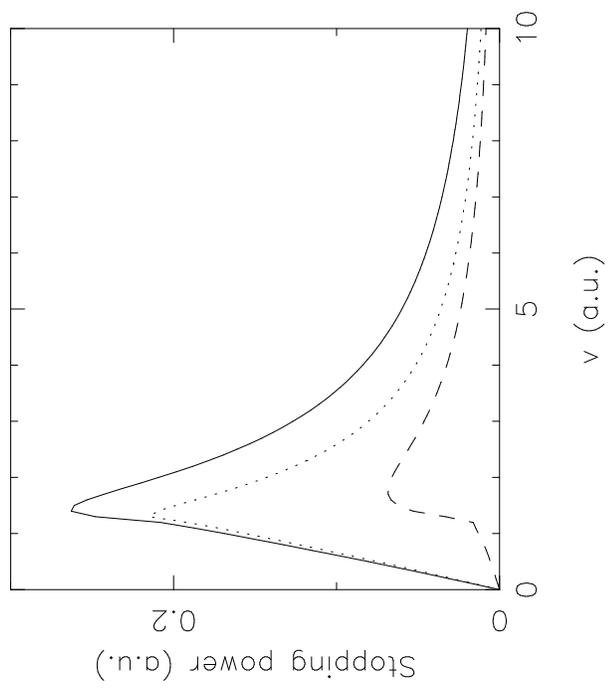
\begin{figure}
\caption{Full RPA $Z_1^2$ contribution to the stopping power (solid line),
versus velocity. Dashed and dotted lines represent total contributions for
$q<q_c$ and $q>q_c$, respectively, $q$ representing the momentum transfer, and
$q_c$, the critical momentum for the plasmon being a well-defined excitation.}
\end{figure}

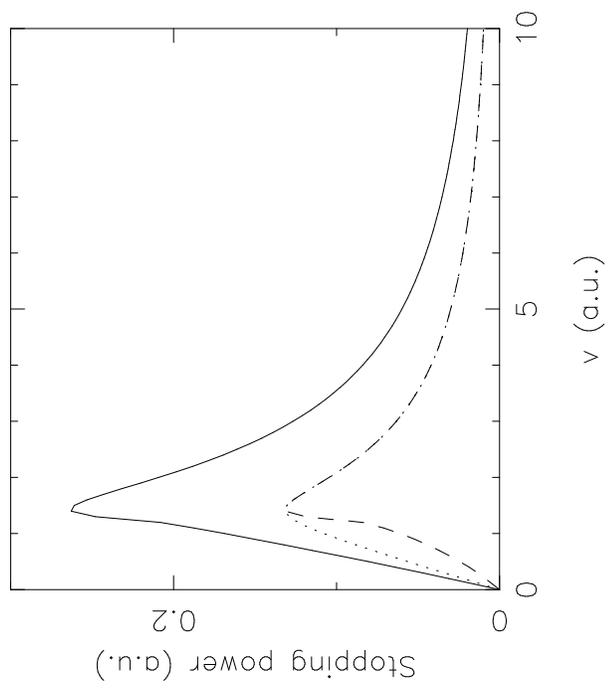
\begin{figure}
\caption{As in Fig. 7, with $q_c$ approximated by its low-density limit:
$q_c=\protect\sqrt\protect{2\omega_p}$.}
\end{figure}

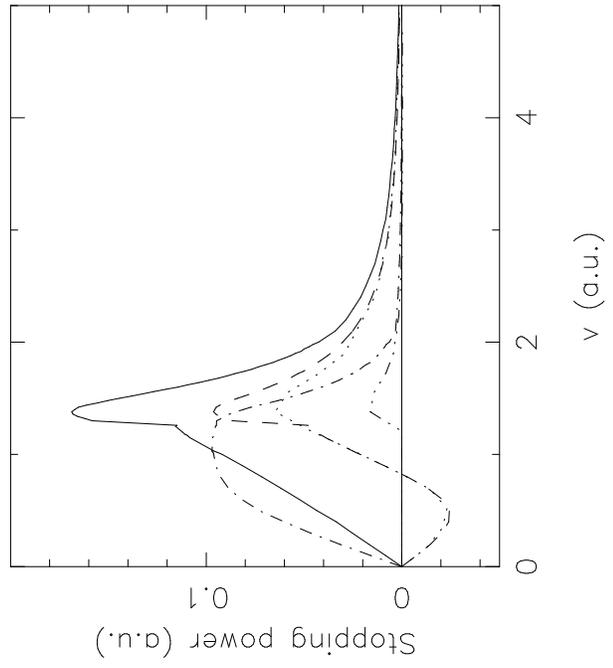
\begin{figure}
\caption{Full RPA $Z_1^3$ contribution to the stopping power (solid line), as
a function of the velocity of the projectile. The total contribution from $f_1$
of Eq. (2.35) (dashed line) has been split into contributions coming from
losses to single plasmons (dashed-dotted-dotted-dotted line) and single
electron-hole pairs (dotted line). The dashed-dotted line represents the total
contribution from
$f_2$ of Eq. (2.36), which appears as a consequence of losses to electron-hole
pair excitations.}
\end{figure}

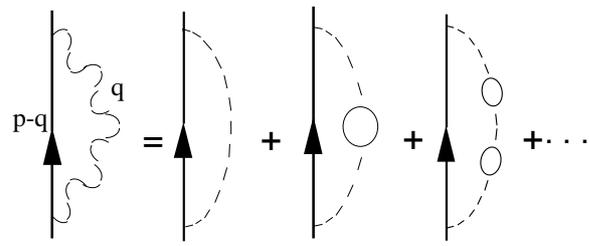
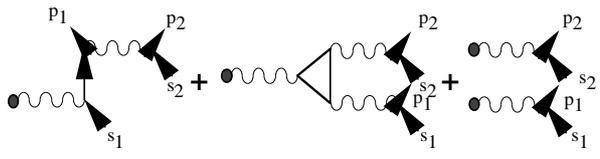
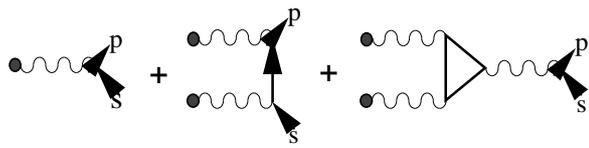
\end{document}